\documentclass{elsart}
\usepackage{graphicx,epsfig,amssymb}
\usepackage{natbib}

\def\astrobj#1{#1}
\journal{New Astronomy}

\begin{document}

\begin{frontmatter}
\title{THE INFLUENCE OF MAGNETIC FIELDS ON THE 
SUNYAEV-ZEL'DOVICH EFFECT IN CLUSTERS OF GALAXIES}

\author{P.\,M.\,Koch\thanksref{cor}},
\thanks[cor]{Corresponding author: pmkoch@physik.unizh.ch}\author{Ph.\,Jetzer}, \author{D.\,Puy}

\address{Institute of Theoretical Physics, University of Z\"urich,
         Winterthurerstrasse 190, CH-8057 Z\"urich, Switzerland}

\begin{abstract}
We study the influence of intracluster large scale magnetic fields on the thermal Sunyaev-Zel'dovich (SZ) effect. In a 
macroscopic approach we complete the hydrostatic equilibrium equation with the magnetic field pressure
component. Comparing the resulting mass distribution with a standard one, we derive a new electron density
profile. For a spherically symmetric cluster model, this new profile can be written as the product of a
standard ($\beta$-) profile and a radius dependent function, close to unity,  
which takes into account the magnetic
field strength. For non-cooling flow clusters we find that the observed magnetic field values can reduce the SZ signal 
by $\sim 10\%$ with respect to the value estimated from X-ray observations and the $\beta$-model. 
If a cluster harbours a cooling flow, magnetic fields tend to 
weaken the cooling flow
influence on the SZ-effect.
\end{abstract}


\end{frontmatter}

\section{Introduction}

The SZ-effect is 
rapidly turning into
an important astrophysical tool thanks to the 
progress 
of the observational techniques, which allow increasingly precise 
measurements. In view of these developments it is thus relevant
to study further corrections to it, such as relativistic effects \citep{re95}, the shape of the
galaxy cluster and its finite extension or a polytropic temperature profile (see e.g. \citet{pu00}), 
corrections induced by halo rotation \citep{co01}, Brillouin scattering \citep{sm02}, early galactic winds \citep{ma01} 
and the presence of cooling flows \citep{sc91,ma01}. These additional effects are of different relevance
and often depend on the specific cluster values.\\

Whereas, e.g. cooling flows are not present in every cluster of galaxies, the need for relativistic corrections due 
to energetic non-thermal electron populations 
seems to be common in most clusters. 
\citep{bl00,sh02}. These relativistic
electrons produce a hard X-ray component in excess of the thermal spectrum by Compton scattering off the Cosmic 
Microwave Background (CMB) and by non-thermal bremsstrahlung. Their emission has been quite possibly
detected 
in \astrobj{Coma} \citep{re99,ff99}, \astrobj{A2199} \citep{ka99}, \astrobj{A2256} \citep{mo00} and \astrobj{A2319} 
\citep{gr02} by RXTE and 
BeppoSAX satellites. The main evidence for the existence of relativistic electrons comes from radio synchrotron 
emission of extended  intracluster regions \citep{gi99,gi00} and is, therefore, 
closely related to the presence of magnetic
fields. The magnetic fields in the intracluster gas  
lead to acceleration processes and modify the classical
Maxwell-Boltzmann distribution of the electrons,
which might acquire a significantly non-thermal spectrum and thus account for the observed
hard X-ray spectra \citep{en99,bl2000}. Consequently, several authors \citep{re95,it98,ch98,bi99} derived relativistic 
corrections to the thermal SZ-effect up to different leading orders. 
Though the magnetic field is related to the 
relativistic electron population, its own influence - as a non-thermal cluster component - on the SZ-effect has not
yet been investigated. Whereas the importance of relativistic corrections to the SZ-effect depends on the cluster 
temperature, the magnetic field seems to be ubiquitous with a mean field value of $5-10\, \mu G$ in the cluster cores
\citep{cl01}. Independently of the diffuse non-thermal radio emission, excess Faraday rotation measure of polarized
radio emission in radio sources within or behind the cluster can prove the existence of magnetic fields. This method 
was applied to
the \astrobj{Coma} cluster, where \citet{fe95} found a large magnetic field of $B\ge 8.3 \,\mu G$. \citet{go99} 
found a similar value ($5-10\, \mu G$) for \astrobj{A119}. \citet{cl99} derived magnetic field strengths of a 
few $\mu G$ for their
cluster sample by using a statistical Faraday rotation measure technique. Unfortunately, the structure of the magnetic
field is presently poorly known. Contrary to the static situation in non-cooling flow clusters, the magnetic field is
believed to become dynamically significant in the cores of cooling flow clusters \citep{ei01}. \citet{ta01} found a 
magnetic field strength of up to $40\,\mu G$ in the \astrobj{Centaurus} cluster. The converging cooling flow 
(see e.g. \citet{fa91,fa94}) causes a compression and enhancement in the magnetic field strength and finally reconnection 
might transfer the magnetic energy back to the
plasma when the magnetic field pressure becomes 
comparable to the thermal gas pressure.\\
As the ratio of the magnetic pressure ($P_B$) to the gas pressure ($P_g$) reaches $\frac{P_B}{P_g}\sim 10^{-2}$ for 
the quoted values of
non-cooling flow clusters and even unity for the cores of cooling flow clusters, we consider the
additional magnetic field pressure term to be significant and we will examine its influence on the SZ-effect. We remark
that all the quoted values refer to large scale magnetic fields with a coherence length of typically $1-10\,kpc$. 
There are essentially no useful limits on the strength of any small scale magnetic fields. 
Our calculation is based on a macroscopic picture inferred from the existence of 
this additional (large scale) pressure component and we do not start our considerations at the level of the single
particle movement in a magnetic field. The current cluster data reveal magnetic field values which result to be 
significant  for a correct SZ analysis, especially towards the cluster core where the electron density increases.\\

The aim of the paper is to examine the influence of large scale magnetic fields on the thermal SZ-effect.
We, therefore, choose a 
phenomenological approach, where 
the addition of a 
magnetic field pressure
to the gas pressure is well justified.
For a given magnetic field model, we can then estimate the change in the electron 
density and the 
temperature profiles as compared to standard ones 
used in the literature in the absence of 
magnetic fields.

The
paper is organised as follows: In section 2 we present the theoretical model for the magnetic field
contribution. We derive new gas density profiles which are then
used to calculate the SZ-effect. We distinguish
two situations according to whether a cluster harbours a cooling flow or not. Section 3 shows our results
and 
contains a
discussion of how magnetic fields influence the SZ signal and what are 
the observational
consequences. As an illustration we apply our results to 
the non-cooling flow cluster \astrobj{A119}. Our conclusions are given in section 4.

\section{Magnetic field contribution}

\subsection{Non-cooling flow clusters}   \label{noncfcluster}

We suppose a standard spherically symmetric
model for a cluster, which is assumed to be in a relaxed state 
with a static gravitational
potential. Thus, hydrostatic equilibrium can be expected.  
The intracluster plasma is treated as an ideal
gas and thus the well known hydrostatic equilibrium equation has the form (see e.g. \citet{sa88}):

\begin{equation}
\frac{1}{\rho_g(r)}\frac{dP_g(r)}{dr}=-\frac{G\,\mathcal{M}(r)}{r^2},\label{ncf1}
\end{equation} 
where $G$ is the gravitational constant, $\rho_g(r)$ and $P_g(r)$ are the radius 
dependent gas density and 
pressure,
respectively, and $\mathcal{M}(r)$ is the gravitating mass within the radius $r$.
$\mathcal{M}(r)$ is mainly determined by the
dark matter profile, whereas the 
gas mass contribution is negligible.
When taking into account magnetic fields, the hydrostatic equilibrium Eq.(\ref{ncf1}) has
to be completed with  
a magnetic hydrostatic pressure term $P_B$ of the form:

\begin{equation}
P_B(r)=\frac{B^2(r)}{8\pi},                     \label{ncf2}
\end{equation}
where $B(r)$ is the magnetic field 
strength, which 
is supposed to be spherically symmetric. Thus, the magnetic field contributes
to the total pressure opposing the gravitational force. 
The gas pressure $P_g(r)$ in 
Eq.(\ref{ncf1}) is replaced by the sum of gas and magnetic field pressure:

\begin{equation}
P_g(r)+P_B(r)=\frac{k}{\mu\,m_p}\rho_g(r)T_g(r)+\frac{B^2(r)}{8\pi},       \label{ncf3}
\end{equation}
where $k$ is the Boltzmann constant, $\mu$ the mean molecular weight, $m_p$ the proton mass
and $T_g(r)$ the gas temperature at radius $r$.\\
When deriving the mass distribution from Eq.(\ref{ncf1}), either with only the gas pressure
or with the magnetic field pressure added as in Eq.(\ref{ncf3}), we obviously end up 
with the two different\footnote[1]{If necessary we label quantities with magnetic fields with an index $'B'$ to 
distinguish them from quantities without magnetic fields.} equations:

\begin{eqnarray}
\mathcal{M}(r)&=&-\frac{k\,T_g(r)}{G\,\mu\,m_p}r\left[\frac{d\ln\rho_g(r)}{d\ln r}
                 +\frac{d\ln T_g(r)}{d\ln r}\right],                        \label{ncf4}\\
\mathcal{M}_B(r)&=&-\frac{k\,T_B(r)}{G\,\mu\,m_p}r\left[\frac{d\ln\rho_B(r)}{d\ln r}
                 +\frac{d\ln T_B(r)}{d\ln r}\right] -
                  \frac{r^2}{G\,\rho_B(r)}\frac{dP_B(r)}{dr},  \label{ncf5}
\end{eqnarray}
where $T_B(r)$ and $\rho_B(r)$ describe the gas temperature and density, respectively, in
the presence of the magnetic field.\\

As next we 
compare the Eqs.(\ref{ncf4}) and (\ref{ncf5}). To find the cluster mass distribution 
we can either proceed our analysis according to Eq.(\ref{ncf4}) or, if we take into account 
magnetic fields, according to Eq.(\ref{ncf5}). 
However, the true total 
gravitating  cluster mass 
is unique and must be the same, thus $\mathcal{M}_B(r_l) 
\equiv\mathcal{M}(r_l)$, 
$r_l$ being the cluster limiting extension.
Nevertheless, $T_B(r)$ and 
$\rho_B(r)$ can be different from $T_g(r)$ and $\rho_g(r)$ due to 
the magnetic field pressure gradient\footnote[2]{We note that our 
starting point is different from the usual observer's point of view: From a data
set, a density profile is first derived and later the magnetic field contribution is 
added as
in Eq.(\ref{ncf5}). This procedure gives a higher 
cluster mass (see e.g. \citet{lm94}).}.
In what follows, we want to relate 
$\rho_B(r)$ to $\rho_g(r)$, which
is supposed to be known and to be a correct theoretical description for the gas
density profile in absence of magnetic fields (e.g. $\beta$-model, \citet{sa88}).\\
As dark matter is the dominant
mass component in clusters of galaxies (up to $80-90 \%$ of the total mass) and supposed not to be affected 
by magnetic fields, we expect in good approximation the equality between the Eqs.(\ref{ncf4}) and 
(\ref{ncf5}) to be satisfied for all radii $r$:

\begin{equation}
\mathcal{M}_B(r)=\mathcal{M}(r).                    \label{ncf6}
\end{equation}
Clearly, this approximation is justified as long as the density 
profiles $\rho_B(r)$ and $\rho_g(r)$ do not differ substantially. As we 
will see, our results do indeed satisfy this requirement.
For the sake of simplicity we assume isothermal temperature profiles: $T_g(r)\equiv T_g$, 
$T_B(r)\equiv T_B$. By setting equal the right hand sides of the Eqs.(\ref{ncf4}) and (\ref{ncf5}), we find a first 
order differential equation for $\rho_B(r)$, which yields:

\begin{equation}
\rho_B(r)=\rho_g(r)^{\frac{T_g}{T_B}}\rho_{g,0}^{1-\frac{T_g}{T_B}}\left[1+\rho_{g,0}^{\frac{T_g}{T_B}-1}
\Big(\frac{\mu\,m_p}{k\,T_B}\Big)\int\limits_r^{r_l}\frac{P'_B(\tilde{r})}{\rho_g(\tilde{r})^
{\frac{T_g}{T_B}}}\,d\tilde{r}\right],                                         \label{ncf7}
\end{equation}
where prime is the derivative with respect to $r$ and $\rho_{g,0}$ is the central gas density.  
The boundary condition is chosen to be $\rho_B(r_l)=\rho_g(r_l)$ for $r_l$ at the cluster limiting
radius, where the magnetic field is negligible. Physically, we do not expect the temperatures
$T_g$ and $T_B$ to differ significantly. Thus, in the sense of a first order development, we suppose
$T_g$ and $T_B$ to be approximately equal, which simplifies the above equation. 
Mathematically, the boundary 
condition and the assumption of isothermal temperature profiles 
even impose $T_B\equiv T_g$: At the limiting radius $r_l$, the magnetic field pressure gradient in
Eq.(\ref{ncf5}) vanishes and the densities are then equal. As the temperatures are isothermal, the second term in 
brackets in the Eqs.(\ref{ncf4}) and (\ref{ncf5}) drops, stating that $T_B=T_g$ at $r_l$, which is then also
true for the whole cluster.\\ 

To evaluate 
Eq.(\ref{ncf7}) we need a magnetic field model. Various physical models have been invoked to explain
the rotation measure in radio sources in clusters of galaxies \citep{ja80,tr91}. According to them, 
the magnetic field distribution is correlated with the electron density of the thermal gas.  
Recently, \citet{do01},
based upon a correlation between X-ray surface brightness and Faraday rotation measure, derived the 
following relation:

\begin{equation}
B(r)\propto (n_e(r))^\gamma,                            \label{ncf8}
\end{equation}
where $n_e(r)$ is the electron number density and $\gamma$ the slope of the $B-n_e$ relation, which depends
on the specific cluster of galaxies. They showed, that Eq.(\ref{ncf8}) can be motivated by both 
simulations and observational data. Furthermore, they clearly excluded the possibility of a constant 
magnetic field through the intracluster medium. With the relation (\ref{ncf8}) the magnetic field
profile is proportional to $\rho_B(r)$ 
and we can thus rewrite Eq.(\ref{ncf7}) as follows:

\begin{equation}
\rho_B(r)=\rho_g(r)\left[1+\frac{1}{\rho_{B,0}^{\,\,2\gamma}}\frac{B_0^{\,2}}{8\pi}
\Big(\frac{\mu\,m_p}{k\,T_g}\Big)\int\limits_r^{r_l}\frac{(\rho_B^{\,\,2\gamma})^{\prime}(\tilde{r})}
{\rho_g(\tilde{r})}
\,d\tilde{r}\right],                                         \label{ncf9}
\end{equation}
where $\rho_{B,0}$ and $B_0$ are the cluster central gas density and the central magnetic field value,
respectively. Eq.(\ref{ncf9}) is an integro-differential equation for $\rho_B(r)$. It
expresses the modified density $\rho_B(r)$ as the standard density $\rho_g(r)$ from Eq.(\ref{ncf4}),
multiplied with a radius dependent function which involves the magnetic field pressure. As in perturbation
theory, $\rho_B(r)$ can also be interpreted as the product of the unperturbed quantity $\rho_g(r)$ 
multiplied by the function in brackets in Eq.(\ref{ncf9}), which is close to unity. 
Following an iterative procedure to solve Eq.(\ref{ncf9}),
$\rho_g(r)$ is reinserted into the function in brackets. We stop after 
the first iteration to get:

\begin{equation}
\rho_B(r)\cong\rho_g(r)\left[1+\frac{1}{\rho_{g,0}^{\,\,2\gamma}}\frac{B_0^{\,2}}{8\pi}
\Big(\frac{\mu\,m_p}{k\,T_g}\Big)\int\limits_r^{r_l}\frac{(\rho_g^{\,\,2\gamma})^{\prime}(\tilde{r})}
{\rho_g(\tilde{r})}
\,d\tilde{r}\right].                                        \label{ncf10}
\end{equation}
Setting $\rho_g(r)=\rho_{g,0}f(r)$, where $f(r)$ is the shape of the gas profile, Eq.(\ref{ncf10}) 
finally reads:      

\begin{equation}
\rho_B(r)=\rho_{g,0}f(r)\left[1+\frac{B_0^{\,2}}{8\pi}
\Big(\frac{1}{k\,T_g\,n_{g,0}}\Big)\int\limits_r^{r_l}\frac{(f^{2\gamma})^{\prime}(\tilde{r})}
{f(\tilde{r})}\,d\tilde{r}\right]\quad < \quad\rho_g(r),                                  \label{ncf11}
\end{equation}
where the last inequality arises immediately because the magnetic field pressure decreases towards the 
cluster boundary. The modified gas density $\rho_B$ can thus be calculated from any standard density $\rho_g$, 
the cluster temperature $T_g$ and some central magnetic field value $B_0$. 
We stress that this result follows 
from the starting point that the cluster mass $\mathcal{M}(r)=\mathcal{M}_B(r)$ can be determined in two different 
ways following the Eqs.(\ref{ncf4}) and (\ref{ncf5}), when the self-gravity of the gas is neglected.

\subsection{Cooling flow clusters}   \label{cfcluster}

We suppose again spherical symmetry. For simplicity, we adopt a homogeneous steady-state cooling flow model. 
The gas has a single temperature and density at a given radius and no mass drops out of the flow. 
The cluster is expected to be in a relaxed state, so 
that hydrostatic equilibrium allows us to use an isothermal $\beta$-model \citep{sa88}.
The dynamics in the cooling flow region can thus be described by a set of Euler equations. Mass, momentum and
energy conservation read \citep{mb78,ws87a,ws87b,sa88}: 

\begin{eqnarray}
& & \frac{1}{r^2}\frac{d}{dr}(r^2\rho_g(r)v(r))=0,                                                 \label{cf1}\\
& &  v(r)\frac{dv(r)}{dr}+\frac{1}{\rho_g(r)}\frac{dP_g(r)}{dr}+\frac{G\mathcal{M}(r)}{r^2}=0,      \label{cf2}\\
& &  v(r)\frac{dE(r)}{dr}-\frac{P_g(r)}{\rho_g^2(r)}v(r)\frac{d\rho_g(r)}{dr}=-\Lambda\rho_g(r).         \label{cf3}
\end{eqnarray}
Here $r$, $v(r)$ and $P_g(r)$ are the radius, gas mean velocity and gas pressure, respectively in the cooling flow. 
The velocity $v(r)$ is defined to be negative for the inward directed cooling flow. The internal energy is 
$E(r)=(3/2)\theta(r)$ with the temperature parameter $\theta$, which defines the square of
the isothermal sound speed $c_s$:

\begin{equation}
\theta(r):=c_s^2(r)= \frac{kT_g(r)}{\mu m_p},
\end{equation}
where $T_g$ is the gas temperature in the cooling flow. $\mathcal{M}(r)$ is the gravitating cluster mass 
inside the  radius $r$. In order to determine it, the above mentioned authors used
the following dark matter ($DM$) density profile:

\begin{equation}
\rho_{DM}(r)=\frac{\rho_0}{1+(r/r_c)^2},
\end{equation}
with a central cluster density $\rho_0=1.8\cdot 10^{-25}g\,cm^{-3}$ and a core radius $r_c=250\,kpc$.
As usual, we assume that the cooling flow makes no significant contribution
to the cluster mass density and that the gas self-gravity can be neglected.\\
The cooling function $\Lambda(\theta)$ is defined so that $\Lambda \rho_g^2$ is the cooling rate per unit volume in 
the gas. We use an analytical fit to the optically thin cooling function \citep{ra76} as given by  
\citep{sw87,ma00}:

\begin{eqnarray}
& & \frac{\Lambda(\theta)}{10^{-22}erg\,cm^3\,s^{-1}}=4.7\cdot\exp\left[-\left(\frac{T}{3.5\cdot10^5K}\right)
^{4.5}\right]\nonumber\\
&+& 0.313\cdot T^{0.08}\cdot\exp\left[-\left(\frac{T}{3.0\cdot10^6K}\right)^{4.4}\right]\nonumber\\
&+& 6.42\cdot T^{-0.2}\cdot\exp\left[-\left(\frac{T}{2.1\cdot10^7K}\right)^{4.0}\right]\nonumber\\
&+& 0.000439\cdot T^{0.35} .                                                       \label{cf4}
\end{eqnarray}
As noted by \citet{ma00}, this fit is accurate to within 4\% 
for a plasma with solar metalicity in the temperature range 
$10^5\le T\le 10^8\,K$. For $10^8\le T\le 10^9\,K$, it underestimates cooling by a factor of order
unity, compared to the exact cooling function as in \citet{st93}.
The continuity Eq.(\ref{cf1}) directly yields an expression for the gas density $\rho_g(r)$ in the cooling flow:

\begin{equation}
\rho_g(r)=\frac{\dot{m}}{4\pi r^2}\frac{1}{v(r)},                       \label{cf4.1}
\end{equation}
where $\dot{m}<0$ is the constant cooling flow mass deposition rate which enters as a parameter in our model.
The Eqs.(\ref{cf1})-(\ref{cf4.1}) describe our standard cooling flow model without magnetic fields.\\

To include the magnetic fields in our calculation we follow the paper by \citet{ss90}.
As it was already argued by them, we also limit our discussion to large scale magnetic fields.
We assume that the magnetic field lines are frozen-in to the inward flowing homogeneous cooling gas.
Under this assumption, the radial and tangential coherence lengths vary as \citep{ss90}:

\begin{equation}
l_r=l_{cool}\left(\frac{v_B(r)}{v_{B,cool}}\right),  \qquad  l_t=l_{cool}\left(\frac{r}{r_{cool}}\right)   ,\label{cf5}
\end{equation}
where $l_{cool}$ and $v_{B,cool}$ are the typical coherence length and inflow velocity, respectively, at the cooling 
radius $r_{cool}\approx 100\,kpc$. The magnetic field is assumed to be isotropic outside of $r_{cool}$, so that 
$l_r=l_t\equiv l_{cool}$ and
$B_r^2=B_t^2/2\equiv B_{cool}^2/3$ for $r\geq r_{cool}$, where $B_{cool}$ 
is the magnetic field strength at 
$r_{cool}$. Inside the cooling flow region the radial and the tangential field components are then:

\begin{equation}
B_r(r)=\sqrt{\frac{1}{3}}B_{cool}\left(\frac{r_{cool}}{r}\right)^2,  \qquad 
B_t(r)=\sqrt{\frac{2}{3}}B_{cool}\left(\frac{r_{cool}}{r}\right)\left(\frac{v_{B,cool}}{v_B(r)}\right).  \label{cf6}
\end{equation}
The compression of the gas is expected to produce a 
sensible increase  
of the frozen-in magnetic field strength.
Thus, the magnetic field lines become increasingly radial as the gas flows inward. As it is pointed out by
\citet{gi02}, this corresponds to the physical condition in the cooling flow with $v_T\ll |v_B|$, where $v_T$ and $v_B$ are 
the turbulent velocity and the mean inflow velocity, respectively. In this model, the turbulence does not disturb the
field geometry during infall. The case $v_T\gg |v_B|$ is considered by \citet{tr93}. 
The field strength in the Soker
and Sarazin model grows faster and in the central region the field reaches values which are an order of magnitude 
larger than the one of Tribble, which becomes more isotropic towards the center. As it is
mentioned by \citet{gi02}, there seems to be observational evidence for Tribble's model in the \astrobj{Perseus} cluster, 
where it was found $v_T\approx 60\,km/s$ and $|v_B|\approx 20\,km/s$ at $r_{cool}$. Contrary to this result, different
authors \citep{cl99,ta99,ta01} quote magnetic field values of the order $B\approx 10-100\,\mu G$ 
for the cores of cooling flow clusters. Furthermore, intense magnetic fields of the order of some tens of $\mu G$
have been derived by detection of extremely high Faraday rotation measures throughout radio galaxies in the centers
of cooling flow clusters \citep{ge93,ta93}.
These results would favour the model by Soker and Sarazin. Being aware of this discrepancy, we
will adopt the Soker and Sarazin model for the following presentation. The magnetic field pressure can then be 
expressed from Eq.(\ref{cf6}):

\begin{equation}
P_B(r)=\frac{P_{B,cool}}{3}\left[\left(\frac{r_{cool}}{r}\right)^4+2\left(\frac{r_{cool}}{r}\right)^2
         \left(\frac{v_{B,cool}}{v_B(r)}\right)^2\right],                 \label{cf7}
\end{equation}
where $P_{B,cool}$ is the magnetic field pressure at the cooling radius $r_{cool}$. From their discussion about
magnetic forces and small scale magnetic fields, \citet{ss90} concluded in adding simply the 
magnetic field pressure term to the gas pressure in the Eqs.(\ref{cf2}) and (\ref{cf3}). \citet{gf99} used then 
the same approach
for their simulation of the evolution of the intracluster medium with magnetic field pressure. \\
Proceeding like this and eliminating the cooling flow gas density $\rho_B(r)$ from the Euler Eqs.(\ref{cf2}) and 
(\ref{cf3}) with Eq.(\ref{cf4.1}), we end up with a system of two coupled first order ordinary differential 
equations for the isothermal sound speed $c_{s,B}(r)=\sqrt{\theta_B(r)}$ and the infall velocity $v_B(r)$ in $r$:

\begin{eqnarray}
\frac{dv_B}{dr} & = &\frac{v_B\left(3G\mathcal{M}-10r\theta_B-\frac{\dot{m}}{2\pi}\frac{\Lambda(\theta)}{v_B^2}-
             16\frac{\pi}{\dot{m}}P_{B,cool}f_1\right)}{r^2(5\theta_B-3v_B^2+8\frac{\pi}{\dot{m}}P_{B,cool}f_3)},  
                                                                                            \label{cf8}\\
\frac{d\theta_B}{dr} & = & 2\left[\frac{\theta_B(2rv_B^2-G\mathcal{M})+\frac{\dot{m}}{4\pi}\frac{\Lambda(\theta)}{v_B^2}
                   \left(v_B^2-\theta_B-\frac{8}{3}\frac{\pi}{\dot{m}}P_{B,cool}f_3\right)+\frac{16}{3}\frac{\pi}
                    {\dot{m}}P_{B,cool}\theta_B f_2}
                    {r^2\left(5\theta_B-3v_B^2+8\frac{\pi}{\dot{m}}P_{B,cool}f_3\right)}\right]\nonumber\\
& & +\frac{8\pi r^2}{\dot{m}}\frac{P_{B,cool}}{3}f_4\left(\frac{2v_B^3r-v_BG\mathcal{M}+\frac{\dot{m}}{6\pi}
\frac{\Lambda(\theta)}{v_B}+\frac{16}{3}\frac{\pi}{\dot{m}}P_{B,cool}v_B f_2}{r^2\left(5\theta_B-3v_B^2+8\frac{\pi}{\dot{m}}
P_{B,cool}f_3\right)}\right).                              \label{cf9}
\end{eqnarray} 

These equations reduce to the system of equations derived
by \citet{mb78} in the absence of magnetic fields, 
$P_{B,cool}\equiv 0$.
The functions $f_i$ are defined as follows:

\begin{eqnarray}
f_1(r)&=&\frac{4}{3}\frac{v_B r_{cool}^4}{r}+\frac{5}{3}\frac{rr_{cool}^2v_{B,cool}^2}{v_B},\\
f_2(r)&=&\frac{v_B r_{cool}^4}{r}-\frac{rr_{cool}^2v_{B,cool}^2}{v_B},\\
f_3(r)&=&\frac{v_B r_{cool}^4}{3r^2}+\frac{8}{3}\frac{r_{cool}^2v_{B,cool}^2}{v_B},\\
f_4(r)&=&\left(\frac{r_{cool}}{r}\right)^4+2\left(\frac{r_{cool}}{r}\right)^2\left(\frac{v_{B,cool}}{v_B}\right)^2.
\end{eqnarray}

Both Eqs.(\ref{cf8}) and (\ref{cf9}) have singularities at the sonic radius $r_s$. Under favorable 
conditions, the sonic singularity corresponds to a crossing of two critical solutions which are asymptotes to
families of hyperbolae near the singularity. This was discussed 
by \citet{mb78} in the absence of 
magnetic fields. The Eqs.(\ref{cf8}) and (\ref{cf9}) allow for transitions from subsonic to supersonic flows, if
the numerators and denominators vanish at $r_s$:

\begin{eqnarray}
0&=&5\theta_B(r_s)-3v_B^2(r_s)+8\frac{\pi}{\dot{m}}P_{B,cool}f_3(r_s),           \label{cf10}\\
0&=&r_s\left(6v_B^2(r_s)-\frac{16\pi}{\dot{m}}P_{B,cool}f_3(r_s)\right)+\frac{\dot{m}\Lambda(T_s)}
    {2\pi v_B^2(r_s)}                                                         \nonumber     \\
 & &+\frac{16\pi}{\dot{m}}P_{B,cool}f_1(r_s)-3G\mathcal{M}(r_s)        \label{cf11},             
\end{eqnarray}
but the quotients are well behaved. Given the form for $\mathcal{M}(r)$, $\Lambda(T)$ and the parameters 
$P_{B,cool}$, $\dot{m}$, $r_{cool}$ and $v_{B,cool}$,  we could solve these equations for possible sonic point(s) 
$r_s$, depending on the chosen temperature $T_s$. This procedure was adopted by several authors who described 
the cluster dark matter density with
a King profile. As it was noted by \citet{mb78} and \citet{su89}, this profile allows 
either one or three solutions for $r_s$ in Eq.(\ref{cf11}) in the absence of magnetic fields. The sonic radius
$r_s$ is then given by the smallest value. To find solutions for the Eqs.(\ref{cf8}) and (\ref{cf9}), in the 
absence of magnetic fields, the above authors started the integration away from the sonic point $r_s$ towards the
cooling radius $r_{cool}$. Since the 
expressions for the derivatives are indeterminate at $r_s$, 
they had to be replaced by nonsingular expressions, derived by making use of 
the Bernoulli-de l'H\^opital's rule, which have to be matched to 
hydrostatic equilibrium outside the cooling 
flow region. As the presence of magnetic fields complicates substantially this procedure, we avoid this 
time-consuming method. For simplicity, we choose a different approach which we describe in section 3.2.
The magnetic field influence in the region outside  
$r_{cool}\approx 100\,kpc$ is calculated by the method described in
section \ref{noncfcluster}.

\section{Results and discussion}

\subsection{Non-cooling flow clusters}

The scope of this section is twofolded: First, we discuss the theoretical impact of the modified density $\rho_B(r)$,
and second, we look at the observational consequences and apply 
as an example our calculations to a specific cluster: 
\astrobj{A119}. Based on the new density profile $\rho_B(r)$, we calculate the change in the SZ-effect.

\subsubsection{Theoretical considerations}   \label{resulttheory}

\begin{figure}
\begin{center}
\includegraphics[scale=1.5]{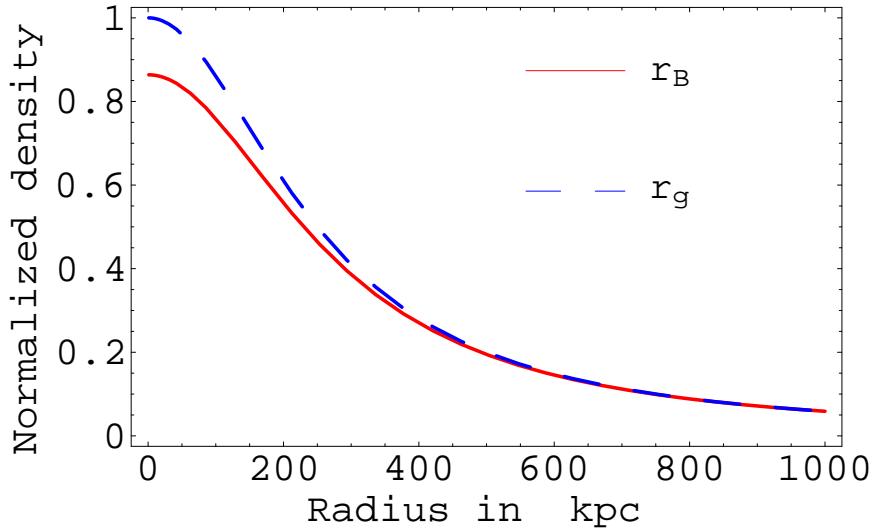}
\caption{\label{graphstandard}The modified profile $\rho_B(r)$ compared to the $\beta$-profile $\rho_g(\beta,r)$
for standard cluster values. The mean molecular weight  
is $\mu=0.63$. The profiles are normalized by
the central gas density $\rho_{g,0}$.}
\end{center}
\end{figure}

\citet{ki66} developed a self-consistent truncated density distribution and gave an analytic function for a 
cluster potential. Based on the hydrostatic equilibrium Eq.(\ref{ncf1}) and a constant temperature, one
derives then the well known isothermal $\beta$-model for the gas density (for a review, see e.g. \citet{sa88}). 
Since then, this model has been extensively used in the literature. In the present discussion we will limit 
the standard input density $\rho_g$ in Eq.(\ref{ncf11}) to this $\beta$-model:

\begin{equation}
\rho_g(\beta,r)=\rho_{g,0}f(r)=\rho_{g,0}\left(1+\left(\frac{r}{r_c}\right)^2\right)^{-3\beta/2},   \label{r1}
\end{equation}
where $r_c$ is the cluster core radius and $\beta$ a fitting parameter. As the $\beta$-model is an appropriate 
theoretically motivated
description in the absence of magnetic fields, Eq.(\ref{ncf11}) states that, observational data, which
involve magnetic fields, should not be fitted with a $\beta$-model. 
The correct fitting function $\rho_B$
is related to a $\beta$-model, $\rho_g(\beta,r)$, through the function in brackets in Eq.(\ref{ncf11}):

\begin{equation}
\rho_B(r)=\rho_g(\beta,r)[1+h_B(r)],                          \label{r1.1}
\end{equation}
where we introduced the notation $h_B(r)$ for the radius dependent magnetic field 
contribution. Eq.(\ref{r1.1}) clearly shows, that the gas density profile
$\rho_B(r)$, as 
used for the surface brightness $S_X$ or energy spectra, is no 
longer of the type of a $\beta$-model.\\
To get an estimate on 
how significantly $\rho_B(r)$ varies compared 
to $\rho_g(\beta,r)$, we adopt some standard
cluster values, which should be free of the magnetic field influence. 
The use of these standard (observational) values will be justified  
in more detail in the next section 
\ref{resultobservation}. The input parameters are: $r_c=250\,kpc$, $T_g=2\cdot10^7\,K$, $n_{g,0}=1.2\cdot10^{-2}cm^{-3}$, 
$\beta=2/3$, $B_0=7.5\,\mu G$ and $\gamma=0.9$. For the isothermal $\beta$-model $\rho_g(\beta,r)$, as introduced
in Eq.(\ref{r1}), the function $h_B(r)$ can be found analytically to yield:

\begin{equation}
h_B(r)=A\cdot\frac{2\gamma}{2\gamma-1}\cdot
     \left\{\left[1+\frac{r_l^2}{r_c^2}\right]^{3\beta(1/2-\gamma)}-
           \left[1+\frac{r^2}{r_c^2}\right]^{3\beta(1/2-\gamma)}\right\},                   \label{r1.2}
\end{equation}
with $A=\frac{B_0^2}{8\pi}\left(\frac{1}{k\,T_g\,n_{g,0}}\right)$, $\gamma\neq 0.5$.\\
Fig. \ref{graphstandard} shows the normalized density profiles for $\rho_g(\beta,r)$ and $\rho_B(r)$. 
We note that the modified profile $\rho_B(r)$
is lower than $\rho_g(\beta,r)$ with the biggest difference of $10-20 \%$ in the most inner part of the cluster,
where the magnetic field becomes important.

Based on these profiles, the SZ-effect can be 
calculated. The frequency dependent intensity change of the cosmic microwave background photons (CMB) due to
inverse Compton scattering off the hot intracluster electrons can be expressed as follows \citep{sz72,re95,bi99}:

\begin{equation}
\Delta I(x)=i_0\, g(x) \int\limits\left(\frac{kT_g}{m_ec^2}\right) \sigma_T n_e\, dl,     \label{r2}
\end{equation}
where $x=\frac{h\nu}{kT}$ is the dimensionless frequency with $T$ the CMB temperature and $i_0=\frac{2(kT)^3}
{(hc)^2}$. The function $g(x)$ defines 
the spectral shape of the thermal SZ-effect.
The integral is the Comptonization parameter $y$ describing the cluster properties with $T_g$,\,$m_e$ 
the electron cluster temperature and electron mass, respectively. $n_e$ is the electron number density in the 
cluster and 
$\sigma_T$ the Thomson cross section. ($k$, $h$, $c$ are the Boltzmann constant, the Planck constant and the 
speed of light, respectively.)
As we are interested in the influence of magnetic fields, we define  $\alpha$ to be the ratio between the SZ-effect 
with the modified density $\rho_B(r)$ and the standard SZ-effect with $\rho_g(\beta,r)$. For our standard values we 
find:

\begin{equation}
\alpha:=\frac{\Delta I_B}{\Delta I}=\frac{\int_0^{r_l}\rho_B(\tilde{r})d\tilde{r}}
        {\int_0^{r_l}\rho_g(\beta,\tilde{r})d\tilde{r}}=0.92,                              \label{r4}
\end{equation}
where the integration is along a line of sight through the cluster center with a limiting radius $r_l=1000\,kpc$. From our theoretical analysis we thus find that magnetic fields reduce 
the SZ-effect by $\approx\,10\%$ compared
to the value one would get for an average standard cluster without magnetic fields.\\
Indeed, $\alpha=\alpha(B_0,T_g,n_{g,0},r_l,\beta,\gamma)$ depends on several observational values.
Modifying the central magnetic field strength $B_0$, while keeping constant the other parameters, can substantially
change the ratio $\alpha$. $B_0\approx 1\,\mu G$ reduces the SZ-effect by less than $1\%$, whereas $B_0\approx 10\,\mu G$ 
gives $\alpha=0.85$. Similarly, a lower temperature $T_g=10^7\,K$ combined with our standard values gives $\alpha=0.84$, 
whereas a very hot cluster with $T_g=10^8\,K$ leads to 
$\alpha=0.98$. In an analogous way, 
$\alpha$ decreases with a smaller 
central gas number density $n_{g,0}$. Reducing $n_{g,0}$ by $20\%$ gives $\alpha=0.90$, instead 
a $20\%$ increase results in 
$\alpha=0.93$. A larger cluster limiting radius $r_l$ makes the magnetic field influence generally more important. 
$\alpha$ is not sensitive to little changes in the parameter $\beta$, because both $\rho_g(\beta,r)$ and $\rho_B(r)$ depend on it.
\citet{do01} estimated the slope $\gamma$ for the $n_e$ - $B$ relation 
to be in the range $\gamma\in (0.5 - 1.0)$. $\alpha$ decreases 
then for
lower values of $\gamma$, reaching $\alpha=0.86$ in the limiting case 
$\gamma=0.5$. All our 
results for $\alpha$, with $\gamma=0.9$, are 
thus still conservative estimates. 

\subsubsection{Observational consequences}   \label{resultobservation}

\begin{figure}
\begin{center}
\includegraphics[scale=1.5]{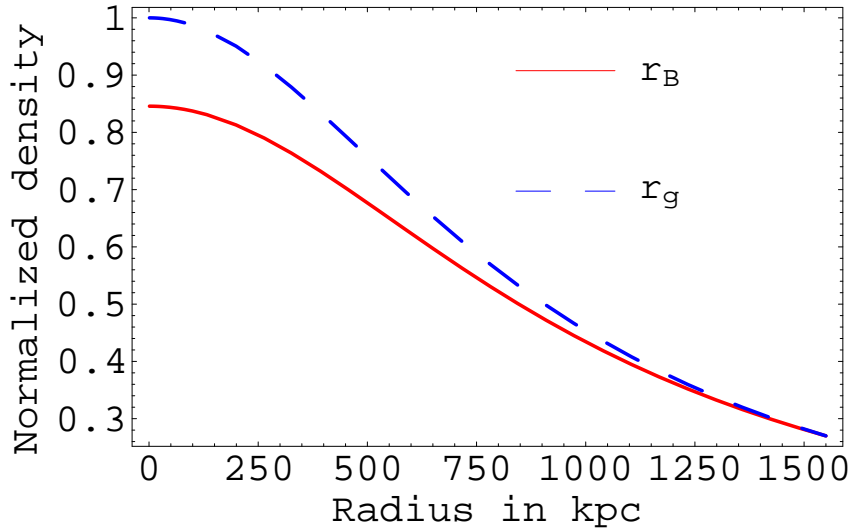}
\caption{\label{graphA119}The modified profile $\rho_B(r)$ compared to the $\beta$-profile $\rho_g(\beta,r)$
for \astrobj{A119}. The mean molecular weight is $\mu=0.63$. The profiles are normalized by the central gas density 
$\rho_{g,0}$.}
\end{center}
\end{figure}

We now switch to an observer's point of view. The observational 
data of a specific cluster, like 
surface brightness and energy spectrum, do already contain the magnetic field effect! As argued before, fitting 
these data with an isothermal $\beta$-model does not 
correspond to the real physical picture. However, the still large error
bars in the 
present data will not allow to discriminate between an isothermal $\beta$-fit and a fit with the profile $\rho_B$.
This is in particular 
true for the value of the central cluster
density, which is often affected by errors of the order of $10 \%$ (see e.g. \citet{mo99}). The isothermal $\beta$-profile 
$\rho_g(\beta_{Fit})$ with the parameter $\beta_{Fit}$,  
determined from observations, must thus be compared
to $\rho_B$:
\begin{equation}
\rho_B=\rho_g(\beta)[1+h_B(r)]\quad \longleftrightarrow \quad \rho_g(\beta_{Fit}).   \label{r5}
\end{equation}  
As the central cluster region is often not very well resolved 
in X-ray data, $\beta_{Fit}$ is usually determined using the data
from the outer parts. In these regions, the magnetic field influence is very 
small, i.e. $h_B\approx 0$. In 
good approximation we get therefore:

\begin{equation}
\beta\approx\beta_{Fit}.                             \label{r6}
\end{equation}
The isothermal $\beta$-model 
will thus overestimate the real density\footnote[3]{These arguments justify the use of the standard cluster values to find 
an estimation for $\rho_B(r)$ in section \ref{resulttheory}.}
$\rho_B(r)$, because $h_B(r)<0$. The error will be largest in the central region where the magnetic field pressure
grows fast and becomes important. The central gas density can be 
overestimated by $10-20 \%$, as it is seen in 
Fig. \ref{graphstandard}.
We, therefore, expect the observed SZ signal to be smaller by $\approx 10\%$ 
compared to the expected value as estimated from X-ray data, when
the magnetic field influence has not been taken into account.\\
Following these arguments, we applied Eq.(\ref{ncf11}) to a well documented non-cooling flow cluster:
\astrobj{A119}, which is characterized by the presence of 
three extended radio galaxies. From different observations remarkably consistent values are found 
(\citet{mo99}, \citet{do01} and references therein) and
the existence of a cooling flow can be excluded for all practical purposes \citep{wh97}. For an isothermal 
$\beta$-model, the cluster parameters\footnote[4]{The cluster 'geometrical' parameters are supposed not to be affected
by the magnetic field.} are: $\beta_{Fit}=0.56$, $\gamma=0.9$, $B_0=7.5\,\mu G$, 
$T_g=5.92\cdot 10^7 K$, $n_{g,0}=2.593\cdot10^{-3}\,cm^{-3}$, $r_c=800\,kpc$, $r_l=1550\,kpc$. From our discussion above, 
the quoted central value $n_{g,0}$ should be overestimated. The other values
are acceptable in 
good approximation, as explained. The results for the normalized 
profiles of $\rho_B(r)$ 
compared to $\rho_g(\beta,r)$ 
are shown in Fig. \ref{graphA119} and the calculated ratio $\alpha$ turns out to be $\alpha\approx 0.905$.
As $n_{g,0}$ is 
overestimated, this is a conservative result and the real influence of the magnetic fields on the SZ-effect
could be even a few percent larger. Reducing the central density 
by $10 \%$ gives a ratio of 
$\alpha=0.895$. If magnetic fields are neglected in \astrobj{A119}, the SZ-effect is thus overestimated by $\approx 10 \%$.

\subsection{Cooling flow clusters}
\citet{ma00} already studied the effect of cooling flows on the SZ distortion and the possible
cosmological implications, in particular on the Hubble constant.
In this section we analyse how magnetic fields, in top of their 
result,
can influence the SZ signal. First, we discuss the solution of the Eqs.(\ref{cf8}) and (\ref{cf9}) with magnetic fields, 
and second, we give the results of our SZ calculation for the magnetic cooling flow model described in section 
\ref{cfcluster}.
We distinguish the cooling flow region from the region outside $r_{cool}$. The later one is analysed 
following the result obtained in section \ref{noncfcluster}. \\

Once the sonic radius $r_s$ is found from the Eqs.(\ref{cf10}) and (\ref{cf11}),
integrating away from $r_s$ requires then to find nonsingular expressions for the derivatives of the differential 
Eqs.(\ref{cf8}) and (\ref{cf9}). This is a non-trivial
task with the additional magnetic field contribution. Furthermore, this procedure (shooting method) requires an
iterative process to match the hydrostatic boundary conditions at $r_{cool}$. If we want to separate the magnetic field
influence, this gets even more complicated as we would have to
match both $\rho_B(r_{cool})$ and $\rho_g(r_{cool})$ with the same cooling flow parameter $\dot{m}$. We would 
thus have to find possible values for $\theta_B(r_s)$ 
and $\theta(r_s)$  at the sonic radius - which differ by 
the magnetic field contribution in the Eqs.(\ref{cf8}) and (\ref{cf9}) -  
which must then lead to the required values 
for $\rho_B(r_{cool})$ and $\rho_g(r_{cool})$ at the cooling radius. 
Since these values at $r_{cool}$ are derived by the procedure
described in section \ref{noncfcluster}, they are not independent from each other neither, but related
through the magnetic field strength $B(r_{cool})$. \citet{sw87}
were already aware of how sensitive the integration of the Eqs.(\ref{cf8}) and (\ref{cf9}), in the absence of
magnetic fields, can be. Not only three boundary conditions ($\rho,\theta,v$) must be matched, but the possible values 
of them are
additionally constrained by three physically motivated conditions like hydrostatic equilibrium, mass deposition
rate and vanishing numerators in the Eqs.(\ref{cf8}) and (\ref{cf9}) for a regular remaining flow. Not 
surprisingly that the range of allowed boundary conditions to find a transonic flow is extremely small.\\
We stress that our goal is not to develop a sophisticated 
cooling flow model, but to get an estimate of how magnetic 
fields can modify the SZ-effect in a cooling flow cluster. We therefore avoid the time consuming and complicated
procedure as outlined above, and start our integration with physically reasonable parameters from $r_{cool}$ towards $r_s$.
The transition at $r_s$ between subsonic and supersonic is always accompanied by shocks. Moreover, this most
inner part will also be under the influence of a central galaxy. Since the interplay between cooling flows, 
central galaxy, shocks and the SZ-effect is not clear, we do not attempt to find solutions to the Eqs.(\ref{cf8}) 
and (\ref{cf9}) inside the sonic radius $r_s$. We will, therefore, 
cut out this supersonic region for our
SZ calculation. Converging cooling flows amplify the magnetic field strength enormously until the magnetic field
pressure ($P_B$) becomes comparable to the thermal gas pressure ($P_g$). Our numerical studies show that this occurs
typically at $\sim 10\,kpc$. The ultimate fate of the magnetic field is still a matter of debate. Field line 
reconnection or convective motions resulting in buoyantly rising regions might be possible answers. Furthermore,
\citet{ch96} pointed out, that more realistic magnetized models can in 
this region not be treated correctly in
spherical symmetry. Based on the above arguments we decided to stop the integration of the Eqs.(\ref{cf8}) and
(\ref{cf9}) at the pressure equipartition radius $r_B (>r_s)$, where $\frac{P_B}{P_g}\sim 1$.\\

We briefly discuss the input values for our results.
The gravitating cluster mass $\mathcal{M}(r)$ is fixed as described in section \ref{cfcluster}. We assume that 
the cooling flow gas makes no significant contribution to the mass density of the cluster.
For the region outside $r_{cool}$, a standard isothermal $\beta$-model with $\beta=\frac{2}{3}$, $T_g=1.9\cdot 10^8\,K$ 
and a central gas density $n_{g,0}=1.2\cdot 10^{-2}\,cm^{-3}$ is consistently adopted. We choose a magnetic field strength 
$B(r_{cool})\approx 1\,\mu G$, which reproduces the observed values 
of the order of $B_0\approx 50\,\mu G$ for the cluster core, 
as found from the cooling flow model in the Eq.(\ref{cf6}). Assuming again $\gamma =0.9$, these values determine then with the 
Eq.(\ref{ncf11}) the 
density profiles $\rho_g(r)$ and $\rho_B(r)$ for the region outside $r_{cool}$, and define the
starting values $\rho_g(r_{cool})$ and $\rho_B(r_{cool})$ for the cooling flow region. Once the mass deposition rate 
$\dot{m}$ is fixed, the continuity Eq.(\ref{cf4.1}) completely determines the initial velocities $v_B(r_{cool})$ and 
$v(r_{cool})$ for the integration of the Eqs.(\ref{cf8}) and (\ref{cf9}). For these velocities at $r_{cool}$ we must 
require that they are of the order of a few tens of $km/s$, which corresponds to the turbulent velocity $v_T$. 
From the difference between $\rho_B(r_{cool})$ and $\rho_g(r_{cool})$, ($\rho_B(r_{cool})\stackrel{<}{\sim}\rho_g(r_{cool})$),
it is obvious that $|v_B(r_{cool})|\stackrel{>}{\sim}|v(r_{cool})|$. The initial sound speed squared, 
$\theta_B(r_{cool})\equiv \theta(r_{cool})$,
is directly related to $T_B\equiv T_g$. Finally, the so derived 
initial conditions and chosen parameters have to allow 
for the existence of a cooling flow: $t_{cool}\stackrel{<}{\sim}t_{Hubble}$. From the above discussion it is clear that 
the integration with possible values from $r_{cool}$ towards $r_s$ is also limited by constraints, which can be summarized 
as follows:

\begin{eqnarray}
& & \mathcal{M}(r)= \mathcal{M}_B(r),        \label{rcf2} \\
& & \rho_g(r_{cool})=\frac{\dot{m}}{4\pi r_{cool}^2}\frac{1}{v(r_{cool})}, \qquad
                        |v(r_{cool})|\simeq v_T,                 \label{rcf3} \\
& & \rho_B(r_{cool})=\frac{\dot{m}}{4\pi r_{cool}^2}\frac{1}{v_B(r_{cool})}, \qquad
                        |v_B(r_{cool})|\simeq v_T, \label{rcf4} \\
& & t_{cool}=\frac{5}{2}\frac{\theta}{\rho\Lambda} \stackrel{<}{\sim} t_{Hubble}\simeq 10^{10}yr.  \label{rcf5}
\end{eqnarray}

These constraints have to be fulfilled with observationally reasonable values $\dot{m}$ and $B_0$. As a possible
set of initial values and cluster parameters we have chosen: $B(r_{cool})=1\,\mu G$, $\dot{m}=-300\,M_{\odot}/yr$, 
$\sqrt{\theta(r_{cool})}\equiv \sqrt{\theta_B(r_{cool})}=1600\,km/s$, $|v(r_{cool})|=|v_B(r_{cool})|=15\,km/s$.\\
\begin{figure}
\includegraphics[scale=0.5]{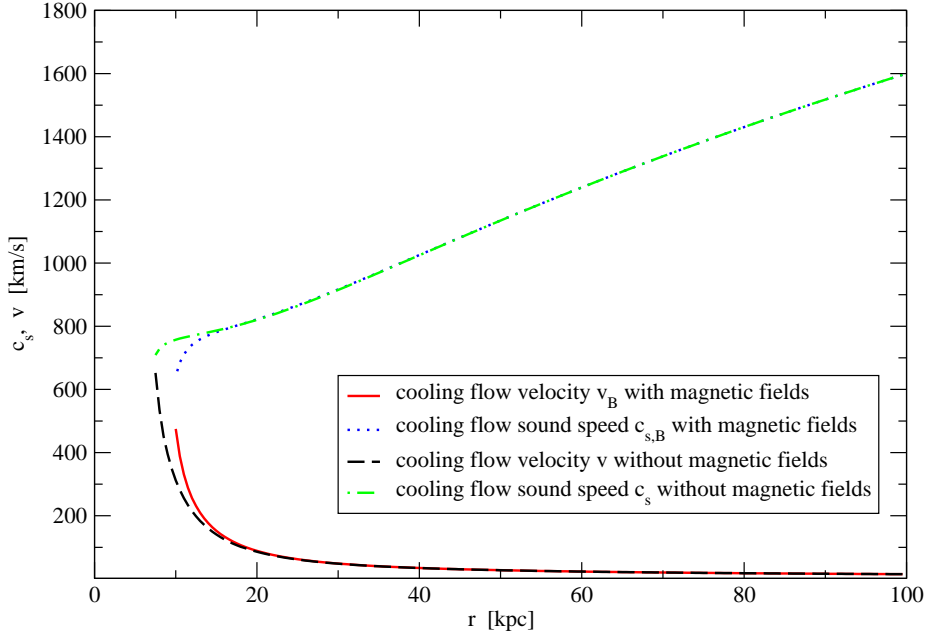}
\caption{\label{cfdynamics}The influence of magnetic fields on the cooling flow velocity $|v(r)|$ and the isothermal
sound speed $c_s(r)$ as a function of radius. The mass deposition rate is $\dot{m}=-300\,M_{\odot}yr^{-1}$, 
$r_{cool}=100\,kpc$ and $B(r_{cool})=1\mu G$. The initial conditions for the integration are:
$\sqrt{\theta(r_{cool})}=\sqrt{\theta_B(r_{cool})}=1600\,km/s$ and $|v(r_{cool})|=|v_B(r_{cool})|=15\,km/s$.}
\end{figure}
Fig.\ref{cfdynamics} shows how the cooling flow dynamics are influenced by magnetic fields. The integration for the profiles
with magnetic fields is stopped at the pressure equipartition radius $r_B$, whereas the profiles in absence of magnetic
fields are stopped close to the sonic radius $r_s$ where the 
Mach number is $M\approx 0.9$.
\begin{figure}
\includegraphics[scale=0.5]{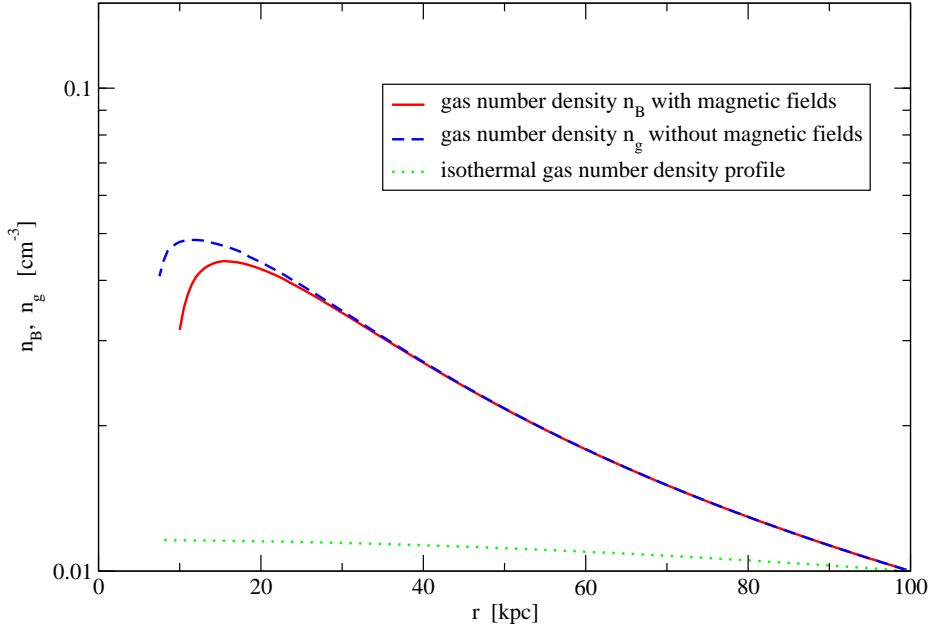}
\caption{\label{cfdensity}The influence of magnetic fields on the cooling flow gas number density. Parameters and 
initial conditions as adopted in Fig.\ref{cfdynamics}.}
\end{figure}
Fig.\ref{cfdensity}
and Fig.\ref{cftemperature} give the corresponding gas number density and temperature profiles for the cooling flow
region. Since the magnetic field strength at $r_{cool}$ is very weak, but the cluster temperature high, the difference 
between $\rho_g(r_{cool})$ and $\rho_B(r_{cool})$ is completely negligible, as it results from Eq.(\ref{ncf11}). 
Therefore, the initial velocities $|v(r_{cool})|$ and $|v_B(r_{cool})|$ can be set equal for all practical 
purposes. 
The influence of the magnetic fields on the SZ-effect is thus entirely determined by the cooling flow region.\\
Similarly to Eq.(\ref{r4}), we define a ratio $\alpha_{CF}$ for the cooling flow $(CF)$ region. For our parameters we
find:
\begin{equation}
\alpha_{CF}:=\frac{\Delta I_{B,CF}}{\Delta I_{CF}}=\frac{\int_{r_B}^{r_{cool}}\rho_B(\tilde{r})T_B(\tilde{r})d\tilde{r}}
        {\int_{r_B}^{r_{cool}}\rho_g(\tilde{r})T_g(\tilde{r})d\tilde{r}}=0.98,                  \label{rcf6}
\end{equation}
where we cut out the most central part of the cluster and the lower integration limit is taken to be the pressure
equipartition radius $r_B=10\,kpc$. 
If the above ratio is completed
with the SZ contribution from the isothermal
region outside $r_{cool}$, ($T_g\int_{r_{cool}}^{r_l}\rho_{iso}(\tilde{r})d\tilde{r},$ with $\rho_{iso}$ an isothermal
$\beta$-model and $r_l=1000 \,kpc$), one finds $\alpha_{CF}\approx 0.99$. The presence of magnetic fields reduces thus only
slightly the cooling flow correction to the SZ-effect.  
When calculating the ratio $\alpha_B$ between the magnetic field influenced profiles
and the standard isothermal profiles, we find instead:
\begin{equation}
\alpha_B:=\frac{\Delta I_{B,CF}}{\Delta I}=\frac{\int_{r_B}^{r_{cool}}\rho_B(\tilde{r})T_B(\tilde{r})d\tilde{r}
                                               +T_g\int_{r_{cool}}^{r_l}\rho_{iso}(\tilde{r})d\tilde{r}}
        {T_g\int_{r_B}^{r_l}\rho_{iso}(\tilde{r})d\tilde{r}}=1.01,                              \label{rcf7}
\end{equation}
compared to $1.02$, which we get without taking into account the magnetic field in the cooling flow.
This is not surprising because of the weakness of the initial magnetic field strength at $r_{cool}$,
which leads to practically identical initial conditions. The magnetic field strength becomes only
important towards the very center and there it then almost compensates the small cooling flow effect.
For the chosen parameters, magnetic field and even cooling flow do not modify much the SZ-effect.\\
Nevertheless, our numerical investigations showed that for a higher magnetic field strength, $B(r_{cool})\approx 6\,\mu G$,
and a higher temperature, $T_g\approx 5\cdot 10^8\,K$, the ratio in Eq.(\ref{rcf6}) can reach $\alpha_{CF}=0.94$.
Moreover, at variance with the above discussed example, a higher magnetic
field strength at $r_{cool}$ leads also to a sizeable lower gas density,
as compared to the case without magnetic field, in the region outside
$r_{cool}$. This fact, as discussed in Section 3.1.1, modifies accordingly
the contribution to the SZ-effect coming from the integration outside $r_{cool}$.
For such cooling flow parameters, \citet{ma00} expect an overestimation of the standard SZ-effect by $\sim 10\%$, due to
the presence of a non-magnetized cooling flow. We instead find that  
this overestimation gets reduced to 
$5-8\%$ when the (higher) magnetic field is taken into account.
Finally we note that our results are conservative estimates since we always
cut out the most central part of the cluster. 

\begin{figure}
\includegraphics[scale=0.5]{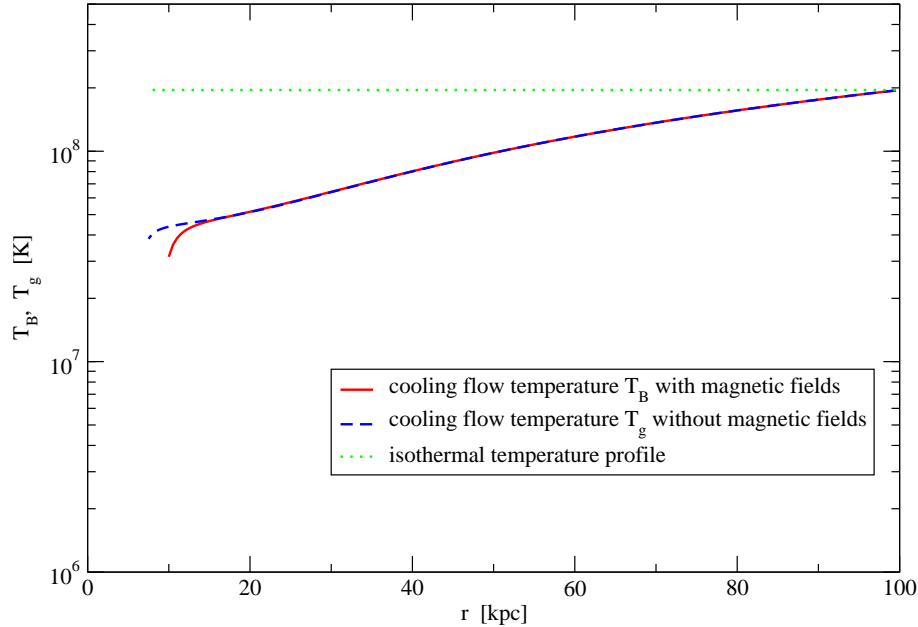}
\caption{\label{cftemperature}The influence of magnetic fields on the cooling flow temperature. Parameters and 
initial conditions as adopted in Fig.\ref{cfdynamics}.}
\end{figure}

\section{Conclusions}

In a phenomenological approach we added the magnetic field pressure to the gas pressure and for clusters without
cooling flows we derived  
in a perturbative procedure a new gas density profile. 
This can be related to a 
standard $\beta$-profile and a function, which takes into account the radius dependent magnetic field strength,
which is assumed to be correlated to the electron density.
For reasonable cluster parameters we find that magnetic fields reduce the standard SZ signal by $\sim 10\%$. 
Indeed, a reduction of up to $\sim 15\%$ seems possible. Our perturbative approach based on the equal mass 
assumption, $\mathcal{M}_B(r)=\mathcal{M}(r)$, 
turns out to be well 
justified by this order of magnitude correction: The 
corresponding decrease in the gas density is $\le 10-15\%$ and, therefore, the change in the dark matter dominated
total cluster mass profile is less than $1\%$. 
The simplifying assumption of an isothermal temperature is adequate,
because the interplay between temperature and a variable magnetic field strength is not yet very clear \citep{do01}.\\
Furthermore, our considerations showed that the central 
cluster gas density is probably overestimated by $10-20\%$
when fitted with a standard $\beta$-model. 
Better data in future might reveal the need for a modified $\beta$-profile as discussed here.\\
Other possible causes of deviations from the (standard) $\beta$-model have been discussed in the literature in 
the context of the determination of the Hubble constant (\citet{bi91,in95} and references therein). Generally, the 
finite extension of a cluster (already adopted in our calculation) lowers the SZ signal by $5-10\%$ compared to 
the case of an infinite isothermal $\beta$-model, and requires then a larger core radius $r_c$ in the $\beta$-model.
Other departures from the standard $\beta$-model include asphericity and small-scale clumping. The most extreme 
variation of geometry of the original spherical model is obtained if the unique axis of the prolate ($Z>1$) or oblate 
($Z<1$) gas distribution is oriented along the line of sight. The SZ signal scales then with the factor $Z$, which is 
the ratio of the length of the unique axis to the major or minor axis, respectively. Typically, $0.5<Z<2$, and 
asphericity introduces thus a variation of a factor $\sim 2$ in the core radius $r_c$. Whereas this effect will be 
averaged out in a large enough cluster sample, all cluster atmospheres will be clumpy to some degree. This might result
in an overestimation of the standard SZ signal at the percentage level, but current simulations are limited in 
resolution and clumpiness might be more important. Whereas all these effects modify the parameters of the standard
$\beta$-model, the inclusion of magnetic fields requires an extended $\beta$-profile, which takes explicitly into
account the magnetic field strength. Unfortunately, in an integrated SZ measurement all the effects will sum up and
they can hardly be separated. Precise future X-ray data might then help to reveal the inner cluster structure.

For the cooling flow clusters, following the treatment by \citet{ss90}, 
we generalized the equations for the sound speed and infall velocity
derived by \citet{mb78} such as to include also a magnetic field.
For typical initial magnetic field values at $r_{cool}$, $P_B/P_g\sim 10^{-4}$, we find an equipartition radius
$r_B\sim 10\,kpc$ and we derive profiles which are in agreement with Soker and Sarazin. Our somewhat larger sonic 
radius $r_s\sim 5\, kpc$ - though not irrealistic as it was found by \citet{su89}, who derived sonic radii of 
$\sim 10-20\,kpc$ for plausible cluster and cooling flow parameters - results probably from different initial conditions.
As the integration for the equations of sound speed and infall velocity is very delicate or even impossible for any
arbitrary combination of parameters, our results can not immediately be generalized. Current discussions about the 
magnetic field model in the cooling flow region
limit further our conclusions. Nevertheless, it turns out
that the gas density in the cooling flow gets somewhat smaller in the
presence of magnetic fields as compared to the case without. This
translates then into a weaker influence of the order of some 
percent of the magnetized cooling flow on the SZ-effect. In special cases magnetic fields might almost compensate
the cooling flow correction to the SZ-effect.\\
Indeed, a precise calculation would require more sophisticated models which are adapted to the specific cluster 
parameters. For all these reasons, cooling flow clusters do not seem to be ideal targets for SZ observations.

\noindent{\bf Acknowledgments}

This work was supported by the Swiss National Science Foundation. 
Part of the work of D.P. has 
been conducted under the auspices of 
the {\it D\raisebox{0.5ex}{r} Tomalla Foundation}.

\end{document}